\documentstyle[12pt]{article}

\begin{document}

\begin{flushright}
Alberta Thy-12-96\\
hep-ph/9605293\\
May, 1996
\end{flushright}
\vspace {0.3in}

\begin{center}
{\LARGE\bf Non-Factorization in   $(B, B_s) \rightarrow P(V) \psi$ Decays}\\
\vspace {0.3in}
{\large A. N. Kamal and F. M. Al-Shamali} \\ \vspace {0.1in}
 \small\em Theoretical Physics Institute and Department of Physics, \\
 \small\em University of Alberta, Edmonton, Alberta T6G 2J1, Canada.
\end{center}
\vspace {0.1in}

\begin{abstract}
Available experimental data on decay rate and polarization are used to
investigate non-factorization contribution to processes of the kind $B
\rightarrow K \psi$, and
$B \rightarrow K^* \psi$ using five theoretical models for the formfactors.
Using the knowledge on non-factorization gained from $B$ decays we study the
processes $B_s \rightarrow (\eta', \eta) \psi$,  and $B_s \rightarrow \phi
\psi$ where experimental data are very limited.
\end{abstract}

\begin{center} (PACS numbers: 13.25.Hw, 14.40.Nd) \end{center}

\section{Introduction}
The $1/N_c$ expansion of hadronic matrix elements has been a very important
approach in the study of  weak non-leptonic decays. The leading terms of this
expansion are factorizable into simpler ones, whereas the next to leading terms
are not completely factorizable.

In the standard approach Fierz transformation and color algebra are used to
transform the non-leading contribution into a factorizable part, which is added
to the leading terms, and a non-factorizable part which is neglected
\cite{ref:Wirbel}. This is called the factorization approximation which is
extensively used and sometimes works well.

It was shown by Gourdin, Kamal and Pham \cite{ref:GKamalP94}  that
factorization approximation in all commonly used models of formfactors could
not account for the longitudinal polarization in $B \rightarrow K^* \psi$,
$\Gamma_L / \Gamma$, and the ratio
$(B \rightarrow K \psi)/(B \rightarrow K^* \psi)$. Subsequently, it was shown
in
\cite{ref:KamalS95.3} that inclusion of non-factorized terms enabled one to
understand both data in all the commonly used models of formfactors.

Our aim in this work is to investigate the non-factorization contribution to
processes of the form
$B \rightarrow K \psi$ and $B \rightarrow K^* \psi$ using the experimental data
available on decay rates and polarization. This is a more thorough analysis
than that presented in \cite{ref:KamalS95.3}. Then, in the light of this
investigation we study the processes $B_s^0 \rightarrow \eta \psi$, $B_s^0
\rightarrow \eta' \psi$ and $B_s^0 \rightarrow \phi \psi$.

The layout of this paper is as follows: Section 2 deals with the definitions
and introduces the five
theoretical models of formfactors we employ. Section 3 contains an analysis of
non-factorization in $B \rightarrow K \psi$ and $B \rightarrow K^* \psi$
decays. $B_s$ decays are discussed in Section 4. We end with a discussion in
section 5.

\section{Aim}
\begin{figure}[ht]

\let\picnaturalsize=N
\def\picsize{2.5in}
\def\picfilename{figureA.eps}
\ifx\nopictures Y\else{\ifx\epsfloaded Y\else\input epsf \fi
\let\epsfloaded=Y
\centerline{\ifx\picnaturalsize N\epsfxsize \picsize\fi
\epsfbox{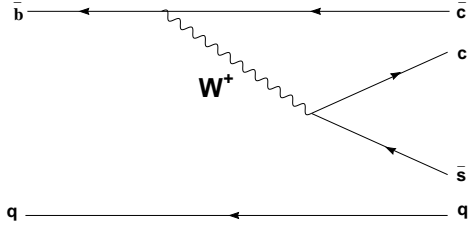}}}\fi

\caption{Quark flow diagram for the two body decay of $B$ meson.}
\label{fig:A}
\end{figure}

In this work we are interested in nonleptonic $B$ decays of the kind
$(B, B_s) \rightarrow P(V) \psi$, where $B$ is $(B^+$ or $B^0)$ , $P$ is
$(K^+, K^0, \eta$ or $ \eta')$ , $V$ is $(K^{*+}, K^{*0}$ or $\phi)$ and $\psi$
represents $\psi$ or $\psi(2S)$. These processes have similar flavor flow
diagrams -- see Fig. \ref{fig:A} -- but different spectator quarks.

\subsection{Decay Rates and Polarization}
Using the effective Hamiltonian that contains the short distance Wilson
coefficients
$C_1$ and $C_2$,  the decay amplitude for such processes is written as
\cite{ref:KamalS95.3}
\begin{eqnarray}
A(B \rightarrow P(V) \psi )  & = & \langle P(V) \psi | {\cal H} ^{\mbox{eff}}_w
| B \rangle \nonumber\\
& = & \frac{G_F}{\sqrt{2}} V^*_{cb} V_{cs} \nonumber\\
& & \times \left[ a_2 \, \langle P(V) \psi | (\overline{b} s) (\overline{c} c)
| B \rangle
+ C_1 \, \langle P(V) \psi |  {\cal H} ^{(8)}_w | B \rangle \right],
\label{eq:Amplitude}
\end{eqnarray}
where
\begin{equation}
{\cal H} ^{(8)}_w = \frac{1}{2} \sum_a (\overline{b} \lambda^a s) (\overline{c}
\lambda^a c),
\label{eq:Non-fac} \end{equation}
and
\begin{equation} a_2 = \frac{C_1}{N_c} + C_2 . \label{eq:Wilson} \end{equation}
$N_c$ is the number of colors and $\lambda^a$ are the Gell-Mann matrices.

At this point, the number of colors,  $N_c = 3$, will be taken seriously. Also,
instead of
neglecting the non-factorizable terms ( second term in (\ref{eq:Amplitude}) and
any non-factorized contribution to the first term ), we paramerize them as in
\cite{ref:KamalS95.3}. The parametrization is done in such a way that we can
conveniently combine the factorizable and the non-factorizable terms.
Explicitly, this means writing the following:

\begin{eqnarray}
\langle \psi |  (\overline{c} c) | 0 \rangle & = & \epsilon^\mu m_\psi f_\psi ,
\label{eq:MED.a} \\
\langle P |  (\overline{b} s) | B \rangle & = & \left( p_B + p_P -
\frac{m^2_B - m^2_P}{q^2} q \right)_\mu F_1(q^2) \nonumber \\
& & + \frac{m^2_B - m^2_P}{q^2} q_\mu F_0(q^2) , \label{eq:MED.b} \\
\langle V |  (\overline{b} s) | B \rangle & = & - \left[  (m_B + m_{V} )
\eta^*_\mu A_1(q^2)
 - \frac{\eta^*.q}{ m_B + m_{V} }  (p_B + p_{V})_\mu A_2(q^2) \right. \nonumber
\\
& & - 2 m_{V} \frac{\eta^*.q}{ q^2} q_\mu (A_3(q^2) - A_0(q^2)) \nonumber \\
& &  \left. - \frac{2 i} { m_B + m_{V}}
\varepsilon_{\mu\nu\rho\sigma} \eta^{*\nu} p^\rho_B p^\sigma_{V} V(q^2) \right]
,
\label{eq:MED.c} \\
\langle P \psi | {\cal H} ^{(8)}_w  | B \rangle & = & 2 m_\psi f_\psi C_P
(\epsilon . p_B)
F^{(8)NF}_1(q^2) , \label{eq:MED.d} \\
\langle V  \psi|  {\cal H} ^{(8)}_w  | B \rangle & = &  - m_\psi f_\psi
 \left[  (m_B + m_{V} ) (\epsilon .\eta^*) A^{(8)NF}_1(q^2)  \right. \nonumber
\\
& &  - \frac{2}{ m_B + m_{V} }  (\epsilon . p_B) (\eta^* . p_B )
A^{(8)NF}_2(q^2) \nonumber \\
& &  \left. - \frac{2 i} { m_B + m_{V}} \varepsilon_{\mu\nu\rho\sigma}
\epsilon^\mu
\eta^{*\nu} p^\rho_B p^\sigma_{V} V^{(8)NF}(q^2) \right] , \label{eq:MED.e}
\end{eqnarray}
where
\begin{equation}
q_\mu = (p_B - p_{P(V)})_\mu = (p_\psi)_\mu . \label{eq:q}
\end{equation}
The polarization vectors $\epsilon^\mu$ and $\eta^\mu$ correspond to the two
vector mesons
$\psi$ and $V$, respectively. In (\ref{eq:MED.d}), $C_P$ arising from the
mixing of $\eta$ and $\eta'$ (see reference \cite{ref:KamalXC94}), is givin by
\begin{equation}
C_P = \left\{
\begin{array}{cl}
\sqrt{\frac{2}{3}} \left( \cos\theta_P + \frac{1}{\sqrt{2}} \sin\theta_P
\right) & P \equiv \eta \\
\sqrt{\frac{2}{3}} \left( \frac{1}{\sqrt{2}} \cos\theta_P -  \sin\theta_P
\right) & P \equiv \eta' \\
1 & P \equiv K^0, K^+
\end{array} \right.
\end{equation}
where the mixing angle is taken to be
\[ \theta_P = - 20^\circ . \]

The non-factorized contribution to the first term in (\ref{eq:Amplitude}) is
parametrized analogously to (\ref{eq:MED.d}) and (\ref{eq:MED.e}) by writing
$F_1^{(1)NF}$, $A_1^{(1)NF}$, etc.\ in place of  $F_1^{(8)NF}$, $A_1^{(8)NF}$,
etc.\ .

Substituting (\ref{eq:MED.a} -- \ref{eq:MED.e}) into the decay amplitude
(\ref{eq:Amplitude}), we can calculate in a straight forward manner decay rates
for the processes $B \rightarrow P(V) \psi$ and polarization for the processes
$B \rightarrow V \psi$. The results are presented below:
 \begin{eqnarray}
\Gamma (B \rightarrow P \psi ) & = & \frac{G^2_F m^5_B}{32 \pi} |V_{cb}|^2
|V_{cs}|^2
a^2_2 \left( \frac{f_\psi}{m_B} \right)^2 |C_P|^2 \nonumber \\
& & \times k^3(t^2) \left| F_1(m^2_\psi) \right|^2
\left| 1 + \frac{C_1}{a_2} \chi_{F1} \right|^2 , \label{eq:GammaBP}\\
\Gamma (B \rightarrow V \psi ) & = & \frac{G^2_F m^5_B}{32 \pi} |V_{cb}|^2
|V_{cs}|^2
a^2_2 \left( \frac{f_\psi}{m_B} \right)^2 \left| A _1(m^2_\psi) \right|^2
\nonumber \\
& & \times k(t^2) t^2 (1 + r)^2 \sum_{\lambda\lambda} H'_{\lambda\lambda} ,
\label{eq:GammaBV}\\
\frac{\Gamma_L }{\Gamma}(B \rightarrow V \psi) & = &
\frac{H'_L}{H'_L + H'_T}, \label{eq:GammaL}
\end{eqnarray}
where
\begin{eqnarray}
H'_L  =  H'_{00} & = & \left[ a \left( 1 + \frac{C_1}{a_2} \chi_{A_1} \right)
- b \left( 1 + \frac{C_1}{a_2} \chi_{A_2} \right) x \right]^2 ,
\label{eq:HL}\\
H'_T  =  H'_{++} + H'_{--} & = & 2 \left[ \left( 1 + \frac{C_1}{a_2} \chi_{A_1}
\right)^2
+ c^2 \left( 1 + \frac{C_1}{a_2} \chi_{V} \right)^2 y^2 \right], \label{eq:HT}
\\
\chi_{F_1} & = & \left( F_1^{(8)NF}(m_\psi^2) +
\frac{a_2}{C_1} F_1^{(1)NF} (m_\psi^2) \right) / F_1 (m_\psi^2) \label{eq:XF1}
\\
\chi_{A_1} & = & \left( A_1^{(8)NF}(m_\psi^2) +
\frac{a_2}{C_1} A_1^{(1)NF} (m_\psi^2) \right) / A_1 (m_\psi^2) \label{eq:XA1}
\\
\chi_{A_2} & = & \left( A_2^{(8)NF}(m_\psi^2) +
\frac{a_2}{C_1} A_2^{(1)NF}(m_\psi^2) \right) / A_2(m_\psi^2) \label{eq:XA2} \\
\chi_{V} & = & \left( V^{(8)NF}(m_\psi^2) +
\frac{a_2}{C_1} V^{(1)NF} (m_\psi^2) \right) / V(m_\psi^2) \label{eq:XV}
\end{eqnarray}
Subscripts $L$ and $T$ in (\ref{eq:HL}) and (\ref{eq:HT}) stand for
'longitudinal' and 'transverse' and $00$, $++$ and $--$ represent the vector
meson helicities.

In (\ref{eq:GammaBP} -- \ref{eq:HT}) we have introduced the following
dimensionless parameters:
\begin{eqnarray}
r & = & \frac{m_{P(V)}}{m_B}, \\
t & = & \frac{m_\psi}{m_B}, \\
k(t^2) & = & \sqrt{ (1 - r^2 - t^2)^2 - 4 r^2 t^2}, \\
a & = & \frac{1 - r^2 - t^2}{2 r t}, \label{eq:a}\\
b & = & \frac{k^2(t^2)}{2 r t (1 + r)^2},  \label{eq:b}\\
c & = & \frac{k(t^2)}{(1 + r)^2}  \label{eq:c}.
\end{eqnarray}
Furthermore, $x$ and $y$ represent the following ratios,
\begin{eqnarray}
x & = & \frac{A_2(m_\psi^2)}{A_1(m_\psi^2)}, \\
y & = & \frac{V(m_\psi^2)}{A_1(m_\psi^2)}.
\end{eqnarray}

\subsection{Formfactors}
Before proceeding to calculate the decay rates and polarization we need the
values of the formfactors. In this work we consider five theoretical models.

The first is the original Bauer-Stech-Wirbel model \cite{ref:BSW85,ref:BSW87}
(called BSW I here) where the formfactors are calculated at zero momentum
transfer and
extrapolated to the desired momentum transfer using a monopole form for all the
formfactors $F_1, A_1, A_2$, and $V$\@.

The second is a modification of the above (called BSW II here). In this model
the
extrapolation from the zero momentum transfer is done using a monopole form for
$A_1$
and a dipole form for $F_1, A_2$, and $V$ \cite{ref:GKamalP94}.

The third is the model of Casalbuoni et al.\ and Deandrea et al.\
\cite{ref:Casalbuoni,ref:Deandrea93}, where the normalization at $q^2 = 0$ is
obtained in a model that combines heavy quark symmetry with chiral symmetry for
light degrees of freedom. We call this (CDDFGN) model. Here, all formfactors
are extrapolated with monopole forms.

The fourth is the work of Altomari and Wolfenstein (AW) \cite{ref:Altomari88}
in which the formfactors are evaluated at maximum momentum transfer based on a
non-relativistic quark model. The formfactors, here, are extrapolated with
monopole forms.

The fifth is the non-relativistic quark model by Isgur, Scora, Grinstein and
Wise (ISGW) \cite{ref:ISGW89}. In this model the formfactors are calculated
using harmonic oscillator wave functions for the particle states. This results
in an exponential dependence in (momentum transfer$)^2$ for all formfactors.

The predicted formfactors in these five models relevant to the processes of
interest are shown in Tables \ref{tab:FormfactorF1} and \ref{tab:FormfactorA1}.

\section{Analysis}
In this section we study the non-factorization contribution to the processes
$B \rightarrow K \psi$ and $B \rightarrow K^* \psi$. This has been done in
\cite{ref:KamalS95.3} but our analysis is more thorough. First, we need to
assign values to the CKM matrix elements, the Wilson coefficients and the decay
constants which we take to be, 
\cite{ref:Particle94,ref:KamalS95.1,ref:KamalS95.3}

\begin{eqnarray}
V_{cs} & = & 0.974, \nonumber\\
V_{cb} & = & 0.04, \nonumber\\
C_1 & = & 1.12 \pm 0.01, \nonumber\\
C_2 & = & - 0.27 \pm 0.03, \\
f_\psi & = & 0.384 \pm 0.014 \mbox{ GeV}, \nonumber\\
f_{\psi(2S)} & = & 0.282 \pm 0.014 \mbox{ GeV}. \nonumber
\end{eqnarray}

\subsection{$B \rightarrow K \psi$ Decays}
The processes considered in this paper are color-suppressed. This implies that
the first term in (\ref{eq:Amplitude}) is proportional to $a_2$ which is rather
small, $0.10 \pm 0.03$. In contrast, the coefficient of the second term in
(\ref{eq:Amplitude}) being $C_1$ is an order of magnitude larger. Hence, as can
be seen from (\ref{eq:GammaBP}), even a 10\% non-factorization in $\chi_{F_1}$
causes the decay rate to increase four times, and with a 20\%
non-factorization, the decay rate becomes nine times larger than the value
predicted in the factorization approximation.

Let us first consider the process $B^+ \rightarrow K^+ \psi$ whose decay rate
is more
precisely measured than of the neutral mode. The decay rate formula for this
process, Eq. (\ref{eq:GammaBP}), can be rearranged and written as
\begin{equation}
\chi_{F_1} = \frac{a_2}{C_1} \left[ \frac{\left\{\Gamma (B^+ \rightarrow K^+
\psi )\right\}^{1/2}}{
(78.422 \times 10^{12} \mbox{ GeV}^{-2} \mbox{ sec}^{-1})^{1/2} \; V_{cs}
V_{cb} a_2 f_\psi
F_1(m_\psi^2)} - 1 \right].
\end{equation}

\begin{figure}[ht]

\let\picnaturalsize=N
\def\picsize{3.4in}
\def\picfilename{figureB.eps}
\ifx\nopictures Y\else{\ifx\epsfloaded Y\else\input epsf \fi
\let\epsfloaded=Y
\centerline{\ifx\picnaturalsize N\epsfxsize \picsize\fi
\epsfbox{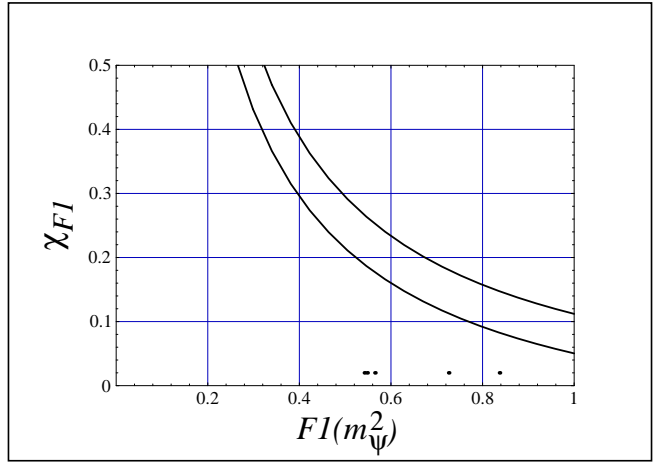}}}\fi

\caption{Allowed region (bounded by the two curves) of
$\chi_{F_1}$ as a function of $F_1(m_\psi^2)$ defined by $B^+ \rightarrow K^+
\psi$. The dots show the model predictions of the formfactors; left: AW, ISGW,
BSW I, CDDFGN, BSW II.}
\label{fig:B}
\end{figure}

By substituting the experimental value for the decay rate \cite{ref:Particle94}
we can calculate the non-factorization contribution as a function of the
formfactor
$F_1(m_\psi^2)$. This is shown in Fig.\ \ref{fig:B}, where the curves represent
$\chi_{F_1}$ with its variance, $\chi_{F_1} \pm \sigma_{\chi_{F_1}}$. The
allowed region lies between the two curves.

This graph shows that if a model predicts a small
value for the formfactor, a high non-factorization contribution is needed in
order to explain the
experimental data. The five models considered in this work predict formfactor
values
in the range (0.55 - 0.84) ( see Table \ref{tab:FormfactorF1}  and the dots in
Fig.\ \ref{fig:B} ) which -- according to Fig.\ \ref{fig:B} -- need about (~10
- 25~\%~) non-factorization.

\begin{figure}[ht]

\let\picnaturalsize=N
\def\picsize{5.0in}
\def\picfilename{figureC.eps}
\ifx\nopictures Y\else{\ifx\epsfloaded Y\else\input epsf \fi
\let\epsfloaded=Y
\centerline{\ifx\picnaturalsize N\epsfxsize \picsize\fi
\epsfbox{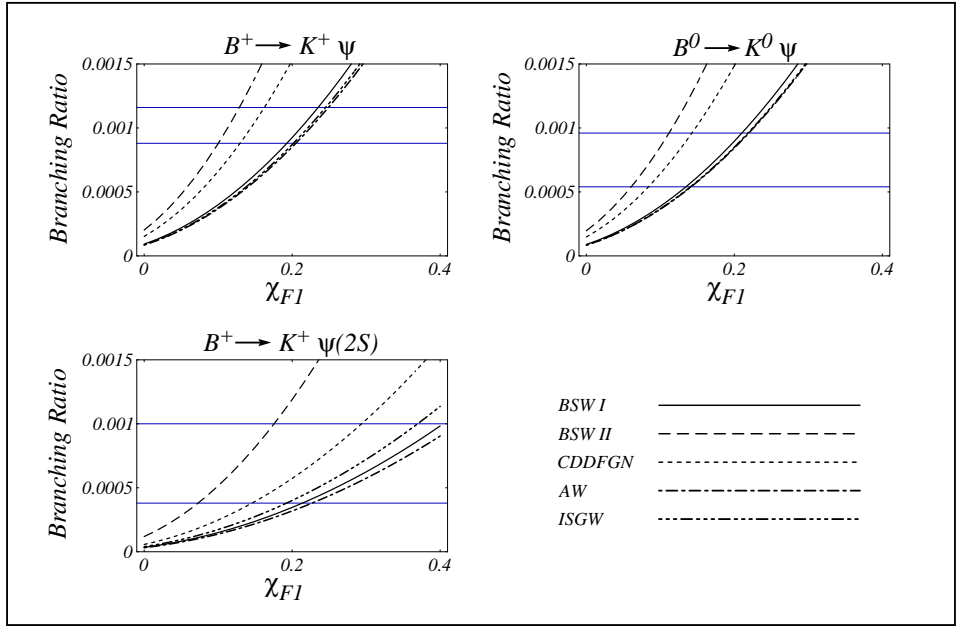}}}\fi

\caption{Branching ratios as a function of $\chi_{F_1}$
predicted by each model. Horizontal lines define the Branching Ratio to one
standard deviation.}
\label{fig:C}
\end{figure}

In a different manner to exhibit the same physics, we show in Fig.\ \ref{fig:C}
the relationship between the branching ratio and the non-factorization
contribution $\chi_{F_1}$ for the processes $B^+ \rightarrow K^+ \psi$, $B^0
\rightarrow K^0 \psi$, and $B^+ \rightarrow K^+ \psi(2S)$ for the five
theoretical models we have considered. The horizontal lines represent one
standard deviation bounds on the branching ratios \cite{ref:Particle94}.

Fig.\ \ref{fig:C} strongly suggests that factorization
($\chi_{F_1} = 0$) does not work, and that non-factorization contribution could
help to explain experimental measurements. The first two graphs in Fig.\
\ref{fig:C}, representing processes $B^+ \rightarrow K^+ \psi$ and $B^0
\rightarrow K^0 \psi$, suggest that one needs (~8 - 25~\%~) non-factorization
contribution to make the model formfactors consistent with data. However, the
third graph, namely that for the process $B^+ \rightarrow K^+ \psi(2S)$,
suggests a relatively higher non-factorization but with poorer precision.
Concerning the decay $B^0 \rightarrow K^0 \psi(2S)$ only an upper limit on the
branching ratio is available, hence we have not plotted the corresponding
graph.

\subsection{$B^0 \rightarrow K^{*0} \psi$ Decay}
Among $B \rightarrow V \psi$, the process $B^0 \rightarrow K^{*0} \psi $ is the
one with the best experimental data. Values for the branching ratio,
$(1.58 \pm 0.28 ) \times 10^{-3}$ \cite{ref:Particle94}, and polarization,
$(0.74 \pm 0.07)$ \cite{ref:Abe95}, are available with reasonable precision for
this exclusive decay. Hence, we concentrate our attention on this neutral mode.

As can be seen from (\ref{eq:GammaBV}) and (\ref{eq:GammaL}), non-factorization
contributions to this decay are parametrized by $\chi_{A_1}, \chi_{A_2},
\mbox{and } \chi_{V}$ which correspond to the three formfactors $A_1, A_2,$ and
$V$. Physics here is more involved. The reason is that we have more unknowns
than we have constraints.

\begin{figure}[ht]

\let\picnaturalsize=N
\def\picsize{3.4in}
\def\picfilename{figureC-2.eps}
\ifx\nopictures Y\else{\ifx\epsfloaded Y\else\input epsf \fi
\let\epsfloaded=Y
\centerline{\ifx\picnaturalsize N\epsfxsize \picsize\fi
\epsfbox{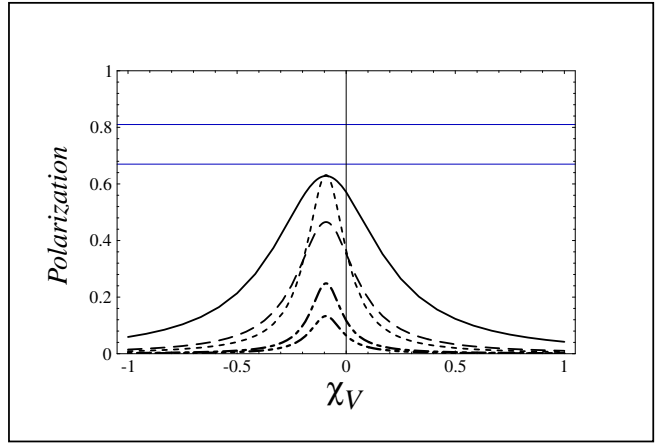}}}\fi

\caption{Polarization for the process $B^0 \rightarrow K^{*0} \psi$ with
$(\chi_{A_1} = \chi_{A_2} = 0)$, plotted as a function of $\chi_V$ for each
model. Horizontal lines define the measured value to one standard deviation.
See Fig.\ ~\ref{fig:C} for legend.}
\label{fig:C-2}
\end{figure}

The simplest approach is to assume that only one formfactor has
non-factorization contribution. We consider this case first. In each of the
five models, we found that allowing only one of  $\chi_{A_1}, \chi_{A_2},
\mbox{and } \chi_{V}$ to be nonzero, one could fit the branching ratio $B(B
\rightarrow K^{*0} \psi)$ with appropriate amounts of non-factorized
contribution. For example, in BSW I model $( \chi_{A_1} = 0.071 \pm 0.024, \:
\chi_{A_2} = 0, \: \chi_V = 0)$ or $( \chi_{A_1} = 0, \: \chi_{A_2} = 0.46 \pm
0.05, \:  \chi_V = 0)$ or $( \chi_{A_1} = 0, \: \chi_{A_2} = 0, \: \chi_V =
0.46 \pm 0.1)$ would fit the branching ratio. However, if we fit the
polarization measurement, we find that there are only two possibilities, $(
\chi_{A_1} = 0.086 \pm 0.098, \:\chi_{A_2} = 0, \: \chi_V = 0)$ and $(
\chi_{A_1} = 0, \: \chi_{A_2} = 0.32 \pm 0.10, \: \chi_V = 0)$ in BSW I model.
There are no solutions to polarization data when $( \chi_{A_1} = 0, \:
\chi_{A_2} = 0, \:  \chi_V \neq 0)$ as shown in Fig.\ \ref{fig:C-2} where we
show that there are no acceptable values of $\chi_V$ that produce large enough
polarization in any of the five models we have considered.

\begin{figure}[ht]

\let\picnaturalsize=N
\def\picsize{5.0in}
\def\picfilename{figureD.eps}
\ifx\nopictures Y\else{\ifx\epsfloaded Y\else\input epsf \fi
\let\epsfloaded=Y
\centerline{\ifx\picnaturalsize N\epsfxsize \picsize\fi
\epsfbox{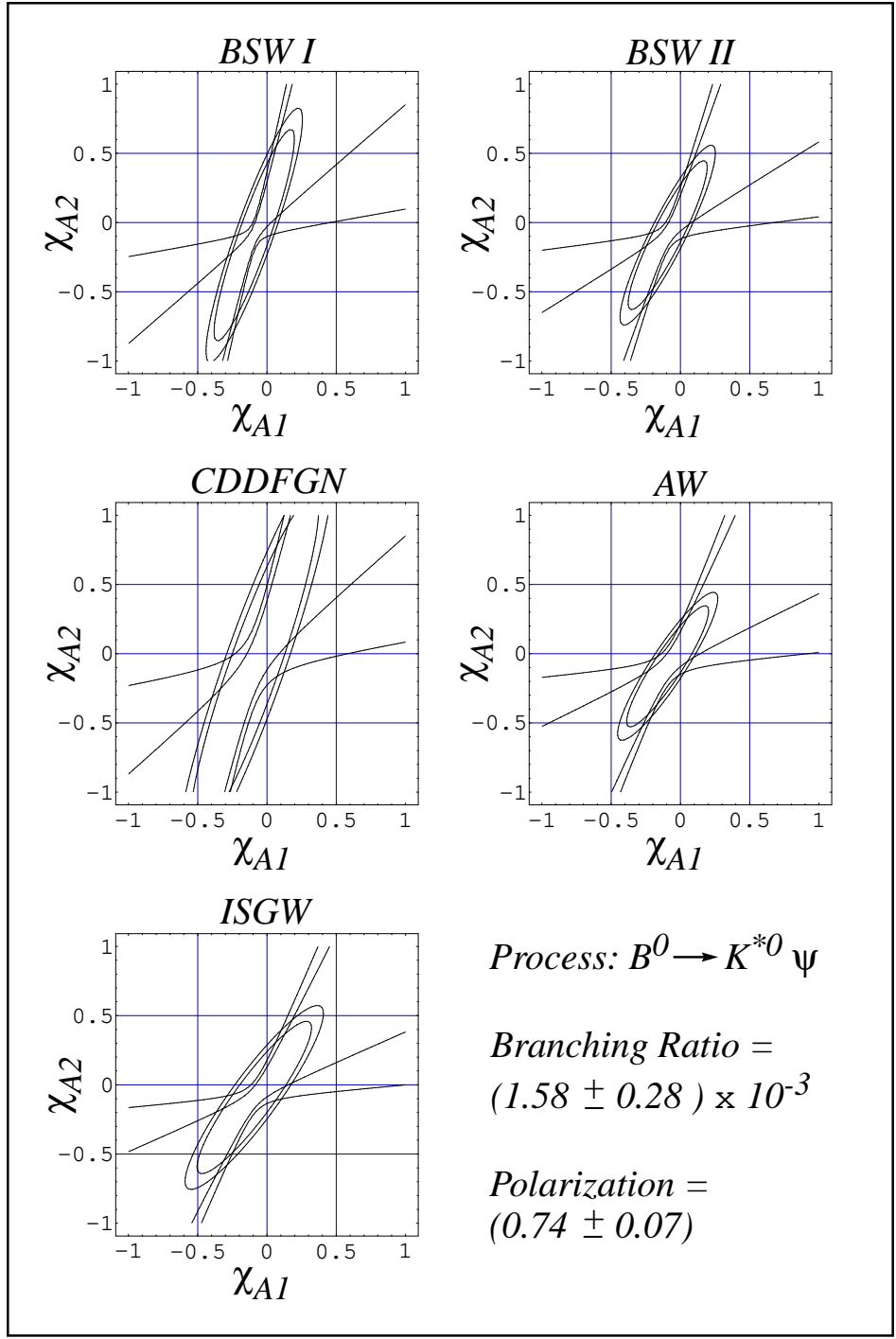}}}\fi

\caption{Regions in $\chi_{A_1} - \chi_{A_2}$ space bounded by
experimental data on branching ratio (ellipses) and polarization (open pairs of
curves) for the process $B^0 \rightarrow K^{*0} \psi$ taking ($\chi_V = 0$).
Values of $\chi_{A_1}$ and $\chi_{A_2}$ allowed by both data lie in the four
areas where the region between the ellipses overlaps with the space between the
two pairs of open curves.}
\label{fig:D}
\end{figure}

Next, we assume that non-factorization is present in both $A_1$ and $A_2$
formfactors while $\chi_V$ is zero. Then the branching ratio demands that the
allowed region in $\chi_{A_1}-\chi_{A_2}$ space lie between two ellipses as
shown in Fig.\ \ref{fig:D} for each of the five models. The region allowed by
the polarization measurement lies, as shown in
Fig.\ \ref{fig:D}, between two pairs of open curves. Thus, in general, there
are four solutions where the domain allowed by polarization overlaps with the
domain allowed by the branching ratio as shown in Fig.\ \ref{fig:D}.

\begin{figure}[ht]

\let\picnaturalsize=N
\def\picsize{3.4in}
\def\picfilename{figureE.eps}
\ifx\nopictures Y\else{\ifx\epsfloaded Y\else\input epsf \fi
\let\epsfloaded=Y
\centerline{\ifx\picnaturalsize N\epsfxsize \picsize\fi
\epsfbox{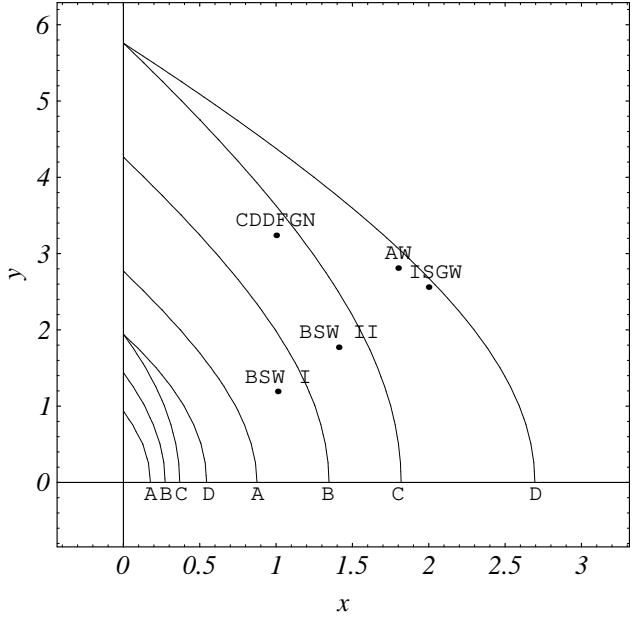}}}\fi

\caption{Regions in $x, y$ plane allowed by experimental data
on polarization for the process $B^0 \rightarrow K^{*0} \psi$. Contours A bound
the region for the case $(\chi_{A_1} = \chi_{A_2} = \chi_{V} = 0)$,
B:$(\chi_{A_1} = 0.05, \chi_{A_2} = \chi_{V} = 0)$, C:$(\chi_{A_1} = 0.1,
\chi_{A_2} = \chi_{V} = 0)$, D:$(\chi_{A_1} = 0.1 , \chi_{A_2} = -0.03,
\chi_{V} = 0)$.The dots represent predictions of the theoretical models.}
\label{fig:E}
\end{figure}

Yet another way to display our results is shown in Fig.\ \ref{fig:E} where we
have plotted the region in $x, y$ plane allowed by polarization data. In the
factorization approximation
$( \chi_{A_1} = 0, \: \chi_{A_2} = 0, \:  \chi_V = 0)$ the allowed region is
bounded by the two curves marked A. If we allow a 5\% non-factorization in
$\chi_{A_1}$ only
(i.\ e.\  $ \chi_{A_1} = 0.05, \: \chi_{A_2} = 0, \:  \chi_V = 0)$, the allowed
region is bounded by the two curves marked B. The region bounded by the curves
C corresponds to
$(\chi_{A_1} = 0.1, \: \chi_{A_2} = 0, \:  \chi_V = 0)$. The case  $(\chi_{A_1}
= 0.1, \: \chi_{A_2} = -0.03, \:  \chi_V = 0)$ gives the region bounded by the
curves D. The values $x$ and $y$ in the five models are also shown in Fig.\
\ref{fig:E}. The values of  $\chi_{A_1}, \chi_{A_2}$, and $\chi_V$ are chosen
only to illustrate the effect of non-factorization on longitudinal
polarization.

\section{$B_s^0$ Decays}
Not much is known about the decays of $B^0_s$ meson. To the best of our
knowledge, only the longitudinal polarization of  $B^0 \rightarrow \phi \psi$
$(0.56 \pm 0.21)$ has been measured \cite{ref:Abe95}, albeit with a large
uncertainty. However, the results of the previous sections regarding the decays
$B^+ \rightarrow K^+ \psi,  B^0 \rightarrow K^{*0} \psi$, \ldots can be used to
study the unknown $B_s^0$ decays if we assume that the amount of
{\em non-factorized contribution is approximately independent of the light
flavor}.

\begin{figure}[ht]

\let\picnaturalsize=N
\def\picsize{5.0in}
\def\picfilename{figureF.eps}
\ifx\nopictures Y\else{\ifx\epsfloaded Y\else\input epsf \fi
\let\epsfloaded=Y
\centerline{\ifx\picnaturalsize N\epsfxsize \picsize\fi
\epsfbox{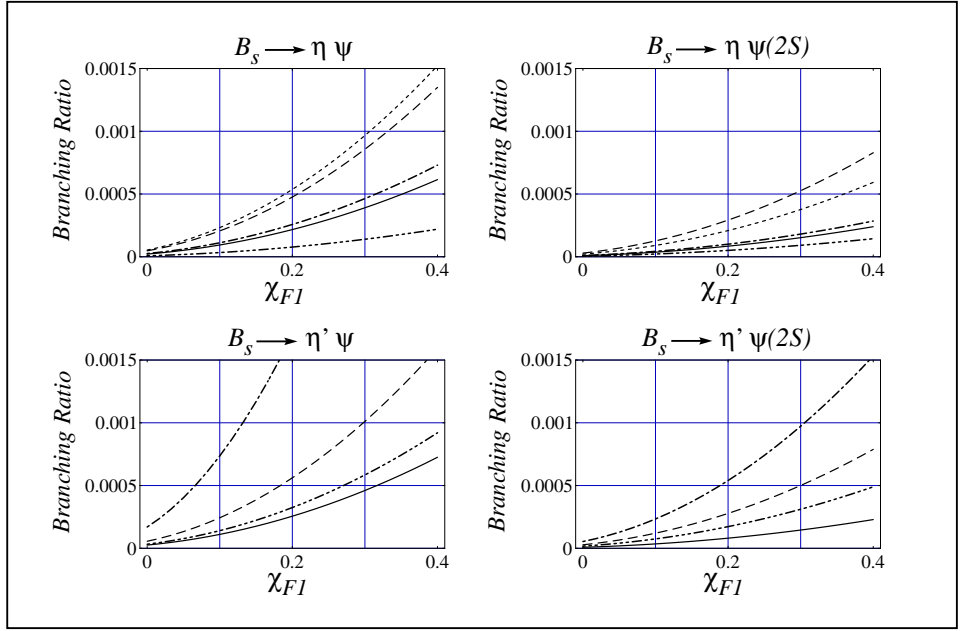}}}\fi

\caption{Branching ratios as a function of $\chi_{F_1}$
predicted by each model. In CDDFGN model, $\eta$ stands for $\eta_8$ and there
is no prediction for $\eta'$. See
Fig.\ ~\ref{fig:C} for the legend.}
\label{fig:F}
\end{figure}

We start with $B_s^0 \rightarrow (\eta, \eta') (\psi, \psi(2S))$ decays. For
these, the branching ratios as a function of $\chi_{F_1}$, are plotted in Fig.\
\ref{fig:F} for each of the five theoretical models. In order to get some feel
for numbers, the branching ratios averaged over the predicted values in the
five models are shown in Table \ref{tab:Bs-eta-psi} for 0~\%, 10~\% and 20~\%
non-factorization contribution.

\begin{figure}[ht]

\let\picnaturalsize=N
\def\picsize{3.4in}
\def\picfilename{figureG.eps}
\ifx\nopictures Y\else{\ifx\epsfloaded Y\else\input epsf \fi
\let\epsfloaded=Y
\centerline{\ifx\picnaturalsize N\epsfxsize \picsize\fi
\epsfbox{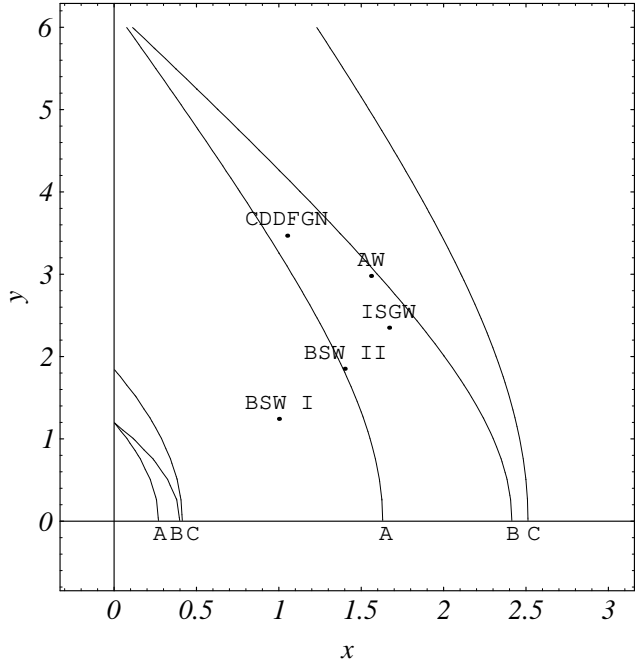}}}\fi

\caption{Regions in  $x, y$ plane allowed by experimental data
on polarization for the process $B^0_s \rightarrow \phi \psi$. Contours A bound
the region for the case  $(\chi_{A_1} = \chi_{A_2} = \chi_{V} = 0)$,
B:$(\chi_{A_1} = 0, \chi_{A_2} = -0.03, \chi_{V} = 0)$, C:$(\chi_{A_1} = 0.05,
\chi_{A_2} = \chi_{V} = 0)$.The dots represent predictions of the theoretical
models.}
\label{fig:G}
\end{figure}

Regarding the process $B^0_s \rightarrow \phi \psi$, we show in Fig.\
\ref{fig:G} the regions in $x, y$ space allowed by the longitudinal
polarization data. The region bounded by contours A corresponds to the
factorization approximation while that bounded by contours C correspond to 5~\%
non-factorization contribution in $\chi_{A_1}$. The region bounded by curves B
corresponds to $(\chi_{A_1} = 0, \: \chi_{A_2} = -0.03, \:  \chi_V = 0)$. This
figure shows that even though the polarization data has low precision,
factorization approximation does not work for all  of the theoretical models
considered.

As we did for the other $B_s^0$ decays in Table \ref{tab:Bs-eta-psi}, we show
in
Table~\ref{tab:Bs-phi-psi} the branching ratios averaged over the five models
and polarization of the processes $B^0_s \rightarrow \phi \psi$ and $B^0_s
\rightarrow \phi \psi(2S)$ for different values of non-factorization
parameters. In Table \ref{tab:BRratio} we have tabulated the ratios
$B(B_s \rightarrow \phi \psi) / B(B^+ \rightarrow K^+ \psi)$ and $B(B_s
\rightarrow \phi \psi) / B(B^0 \rightarrow K^{*0} \psi)$ for different amount
of non-factorization approximation. In this last table the values of  $B(B_s
\rightarrow \phi \psi) $ were taken form Table \ref{tab:Bs-phi-psi} wheras
$B(B^+ \rightarrow K^+ \psi)$ and $B(B^0 \rightarrow K^{*0} \psi)$ were given
the experimental values \cite{ref:Particle94}.

\section{Discussion}
Due to the fact that only two measurements, $B(B \rightarrow K^* \psi)$ and
$P_L(B \rightarrow K^* \psi)$ are available, one can at best derive a
constraint between any two of the three non-factorization parameters
$\chi_{A_1}$, $\chi_{A_2}$, and $\chi_{V}$. The values of these parameters
depend on the theoretical values of the formfactors $A_1$, $A_2$ and $V$. Due
to the fact that of the three parameters, $a$, $b$ and $c$ ( eq.(\ref {eq:a})
-- (\ref {eq:c}) ), $a$ is the largest, it is most economical (largest effect
for the least amount) to put non-factorization in $\chi_{A_1}$. We have shown
that (5 - 10)\% non-factorization in $\chi_{A_1}$ allows the model formfactors
to be consistent with the branching ratio and longitudinal polarization data.

As for the decay $B \rightarrow K \psi$, the five theoretical models we have
chosen require anything from 8\% to 25\% non-factorization contribution
depending on the model.

Assuming that the amount of non-factorized contribution in $(B^0, B^+)$ decays
is independent of the light flavor, we have studied the decays $B_s \rightarrow
(\eta, \eta') \psi$,
$(\eta, \eta') \psi(2S)$, $\phi \psi$ and $\phi \psi(2S)$. To the best of our
knowledge,
only the longitudinal polarization in $B_s \rightarrow \phi \psi$ has been
measured
\cite{ref:Abe95}, though with large errors. This data is consistent with
factorization in
BSW I and BSW II models though not with the other three models we have chosen.
A higher statistics measurement would better test factorization in theoretical
models.

The branching ratios for $B_s \rightarrow (\eta, \eta') \psi$, $B_s \rightarrow
(\eta, \eta') \psi(2S)$, $B_s \rightarrow \phi \psi$ and $B_s \rightarrow \phi
\psi(2S)$ are also quite
sensitive to non-factorization. A measurement of these branching ratios will be
very welcome in testing our ideas.

\vspace{0.20in}
A.N.K thanks the Natural Science and Engineering Research Council of Canada
for the award of a research grant which partially supported this work. F.M.A-S
thanks the
University of Alberta for the award of a Ph.D scholarship which supported this
work.

\newpage
\begin{table}[ht]
\centering
\caption{Model predictions of formfactor $F_1(q^2)$ at $q^2 = m_\psi^2$ or
$m_{\psi(2S)}^2$. In CDDFGN model, $\eta$ stands for $\eta_8$, the octet
member. This scheme cannot handle $\eta_1$, the flavor singlet.}

\begin{tabular}{|l|ccccc|} \hline
& BSW I & BSW II & CDDFGN & AW & ISGW \\ \hline
$B^+ \rightarrow K^+ \psi$ & 0.565 & 0.837 & 0.726 & 0.542 & 0.548 \\
$B^+ \rightarrow K^+ \psi(2S)$ & 0.707 & 1.31 & 0.909 & 0.678 & 0.760 \\
$B^0_s \rightarrow \eta \psi$ & 0.49 & 0.726 & 0.771 & 0.534 & 0.293 \\
$B^0_s \rightarrow \eta \psi(2S)$ & 0.613 & 1.41 & 0.964 & 0.668 & 0.475 \\
$B^0_s \rightarrow \eta' \psi$ & 0.411 & 0.609 & --- & 1.06 & 0.463 \\
$B^0_s \rightarrow \eta' \psi(2S)$ & 0.514 & 0.954 & --- & 1.33 & 0.752 \\
\hline
\end{tabular}

\label{tab:FormfactorF1}
\end{table}
\begin{table}[ht]
\centering
\caption{Model predictions of $A_1(m_\psi^2), A_2(m_\psi^2),$ and $V(m_\psi^2)$
formfactors}

\begin{tabular}{|l|l|ccc|cc|} \hline
\multicolumn{2}{|c|}{} & $A_1$ & $A_2$ & $V$ & $x$ & $y$ \\ \hline
& BSW I &0.458 & 0.462 & 0.548 & 1.01 & 1.19 \\
& BSW II & 0.458 & 0.645 & 0.812 & 1.41 & 1.77 \\
$B^0 \rightarrow K^{*0} \psi$ & CDDFGN & 0.279 & 0.279 & 0.904 & 1.00 & 3.24 \\
& AW & 0.425 & 0.766 & 1.19 & 1.80 & 2.81 \\
& ISGW & 0.316 & 0.631 & 0.807 & 2.00 & 2.56 \\ \hline
& BSW I &0.374 & 0.375 & 0.466 & 1.00 & 1.24 \\
& BSW II & 0.374 & 0.523 & 0.691 & 1.40 & 1.85 \\
$B^0_s \rightarrow \phi \psi$ & CDDFGN & 0.265 & 0.279 & 0.919 & 1.05 & 3.47 \\
& AW & 0.449 & 0.703 & 1.34 & 1.56 & 2.98 \\
& ISGW & 0.237 & 0.396 & 0.558 & 1.67 & 2.35 \\
\hline
\end{tabular}

\label{tab:FormfactorA1}
\end{table}
\pagebreak
\begin{table}[ht]
\centering
\caption{Average branching ratios predicted by the theoretical models for three
choices of
$\chi_{F_1}$.}

\begin{tabular}{|l|c|c|c|} \hline
 & Factorization &  $\chi_{F_1} = 0.1$ &  $\chi_{F_1} = 0.2$ \\
 & $\times10^{-4}$ & $\times10^{-4}$ & $\times10^{-4}$ \\ \hline\hline
$B(B^0_s \rightarrow \eta \psi$) &  0.31 & 1.41 & 3.1 \\
$B(B^0_s \rightarrow \eta \psi(2S)$) &  0.15 & 0.64 & 1.5 \\
$B(B^0_s \rightarrow \eta' \psi$) & 0.71 & 3.1 & 7.1 \\
$B(B^0_s \rightarrow \eta' \psi(2S)$) &  0.27 & 1.2 & 2.7 \\ \hline
\end{tabular}

\label{tab:Bs-eta-psi}
\end{table}
\begin{table}[ht]
\centering
\caption{Average branching ratios and polarization predicted by the theoretical
models for the processes $B^0_s \rightarrow \phi \psi$ and $B^0_s \rightarrow
\phi \psi(2S)$.}

\begin{tabular}{|l|l|c|c|c|c|} \hline
\multicolumn{2}{|c|}{} & Factorization & $\chi_{A_1} =$ 0.05 & $\chi_{A_1} =$
0.1 & $\chi_{A_1} =$ 0.1 \\
\multicolumn{2}{|c|}{} & & $\chi_{A_2} =$ 0 & $\chi_{A_2} =$ 0 & $\chi_{A_2} =$
-0.05 \\
 \hline\hline
$B^0_s \rightarrow \phi \psi$ &  BR & $0.21 \times 10^{-3}$& $0.54 \times
10^{-3}$ &
$1.1 \times 10^{-3}$& $1.4 \times 10^{-3}$\\
 & Pol. & 0.32 & 0.55 & 0.65 & 0.72 \\ \hline
$B^0_s \rightarrow \phi \psi(2S)$ & BR & $0.14 \times 10^{-3}$& $ 0.32 \times
10^{-3}$ &
$0.6 \times 10^{-3}$& $0.67 \times 10^{-3}$ \\
 & Pol. & 0.32 & 0.46 & 0.52 & 0.57 \\ \hline
\end{tabular}

\label{tab:Bs-phi-psi}
\end{table}
\begin{table}[ht]
\centering
\caption{Average branching ratios predicted by the theoretical models for the
process $B^0_s \rightarrow \phi \psi$ divided by the central values of the
experimental measurements of
$B(B^+ \rightarrow K^+ \psi)$ and $B(B^0 \rightarrow K^{*0} \psi)$ for some
chosen values of non-factorization parameters.}

\begin{tabular}{|l|c|c|c|c|} \hline
 & Factorization & $\chi_{A_1} =$ 0.05 & $\chi_{A_1} =$ 0.1 & $\chi_{A_1} =$
0.1 \\
 & & $\chi_{A_2} =$ 0 & $\chi_{A_2} =$ 0 & $\chi_{A_2} =$ -0.05 \\
 \hline\hline
$\frac{B(B^0_s \rightarrow \phi \psi)}{B(B^+ \rightarrow K^+ \psi)}$
& 0.21 & 0.54 &1.1&1.4 \\ \hline
$\frac{B(B^0_s \rightarrow \phi \psi)}{B(B^0 \rightarrow K^{*0} \psi)}$
& 0.13& 0.34 & 0.70 & 0.89 \\ \hline
\end{tabular}

\label{tab:BRratio}
\end{table}
\pagebreak

\end{document}